\documentclass[preprint,12pt]{elsarticle}

\usepackage{amssymb}
\usepackage{amsmath}
\usepackage{algorithm}
\usepackage{algorithmic}
\usepackage{braket}
\usepackage{subfigure}
\usepackage{hyperref}
\usepackage{booktabs}
\usepackage{multirow}
\usepackage{graphicx}
\usepackage{color}

\journal{Neural Networks}

\begin{document}

\begin{frontmatter}



\title{Quantum Mixed-State Self-Attention Network}


\author[label1,label2]{Fu Chen} 
\author[label1]{Qinglin Zhao\corref{cor1}}
\ead{qlzhao@must.edu.mo}
\author[label1]{Li Feng}
\author[label1]{Chuangtao Chen}
\author[label3]{Yangbin Lin}
\author[label4]{Jianhong Lin}
\cortext[cor1]{Co-corresponding author}
\affiliation[label1]{organization={Faculty of Innovation Engineering, Macau University of Science and Technology},
            city={Macau},
            postcode={999078},
            country={China}}
\affiliation[label2]{organization={New Engineering Industry College, Putian University},
            city={Putian},
            postcode={351100}, 
            country={China}}
\affiliation[label3]{organization={Computer Engineering College, Jimei University},
            city={Xiamen},
            postcode={361021}, 
            country={China}}
\affiliation[label4]{organization={Mechanical, Electrical and Information Engineering College, Putian University},
            city={Putian},
            postcode={351100}, 
            country={China}}

\begin{abstract}
Attention mechanisms have revolutionized natural language processing. Combining them with quantum computing aims to further advance this technology. This paper introduces a novel Quantum Mixed-State Self-Attention Network (QMSAN) for natural language processing tasks. Our model leverages quantum computing principles to enhance the effectiveness of self-attention mechanisms. QMSAN uses a quantum attention mechanism based on mixed state, allowing for direct similarity estimation between queries and keys in the quantum domain. This approach leads to more effective attention coefficient calculations. We also propose an innovative quantum positional encoding scheme, implemented through fixed quantum gates within the circuit, improving the model's ability to capture sequence information without additional qubit resources. In numerical experiments of text classification tasks on public datasets, QMSAN outperforms Quantum Self-Attention Neural Network (QSANN). Furthermore, we demonstrate QMSAN's robustness in different quantum noise environments, highlighting its potential for near-term quantum devices. 
\end{abstract}




\end{frontmatter}



\section{Introduction}
\label{sec1}

In recent years, attention-based large language models (such as GPT-4 \cite{radford2018improving,achiam2023gpt} and Claude \cite{enis2024llm}) have achieved remarkable success, significantly excelling in various natural language processing (NLP) tasks. These models can generate complex text \cite{qiu2020pre, min2023recent}, and demonstrate creativity in music composition \cite{haleem2022era}, literary works \cite{min2023recent}, and software development \cite{liu2020multi}, driving rapid advancements in the field of artificial intelligence. While attention mechanisms have demonstrated considerable success on large datasets, they still face certain challenges \cite{zeng2023transformers,hahn2020theoretical}. For example, they may not generalize as effectively on many small to medium-sized datasets \cite{guo2019low}. This limitation arises because, compared to architectures like CNNs and RNNs, attention-based models often require extensive training to infer underlying modality-specific rules, as they do not inherently encode inductive biases suited to particular types of data \cite{khan2022transformers}.

Quantum machine learning (QML), by integrating the strengths of quantum computing with classical machine learning models, offers innovative solutions to the aforementioned challenges \cite{rao2024hybrid,chen2024design}. One distinct advantage of Quantum Machine Learning (QML) is its ability to efficiently map classical data into high-dimensional Hilbert spaces, where the dimensionality grows exponentially with the number of qubits \cite{nielsen2010quantum}. Specifically, an $n$-qubit quantum system can represent a Hilbert space of dimension $2^n$, enabling the processing and analysis of extremely complex data patterns \cite{schuld2019quantum}. This vast quantum space offers new approaches for identifying intricate patterns and improving accuracy in challenging classification tasks when combined with the unique quantum properties of entanglement and superposition \cite{liu2021rigorous,lloyd2021quantum}.
Moreover, quantum algorithms demonstrate significant potential for computational speedup. Shor's algorithm \cite{shor1999polynomial} exploits quantum superposition to perform parallel factorization, while the HHL algorithm \cite{harrow2009quantum} leverages superposition to solve linear systems with exponentially lower time complexity compared to classical methods.

Quantum Natural Language Processing (QNLP) combines the strengths of quantum computing and classical NLP models, offering new possibilities for language processing. However, early models such as Quantum Recurrent Neural Networks (QRNN) \cite{li2023quantum} and Quantum Long Short-Term Memory networks (QLSTM) \cite{chen2022quantum} still struggle to effectively capture long-range dependencies in sequence data.

To address these limitations, researchers began exploring the integration of quantum computing with classical attention mechanisms. In 2022, Zhao et al. proposed the Quantum Self-Attention Network (QSAN) model \cite{zhao2022qsan}, introducing Quantum Logic Similarity (QLS) and Quantum Bit Self-Attention Score Matrix (QBSASM), effectively enhancing the model's ability to extract relevant information. In 2023, Zhao et al. further proposed the Quantum Kernel Self-Attention Network (QKSAN) model \cite{zhao2023qksan}, combining the data representation advantages of quantum kernel methods with the efficient information extraction capability of self-attention mechanisms, providing a larger and more complex data representation space. In 2024,  the Baidu team proposed the Quantum Self-Attention Neural Network (QSANN) model \cite{li2024quantum}, utilizing Gaussian projection quantum self-attention for text classification, capable of exploring word associations in high-dimensional quantum feature spaces. However, these models still have some limitations. For Example, QSANN convert quantum queries and keys to classical data when processing them \cite{li2024quantum}, resulting in information loss. QSAN and QKSAN are limited to pure states \cite{zhao2022qsan,zhao2023qksan}, relying on unitary transformations of quantum circuits, which restricts their expressive power. Additionally, most of the currently implemented quantum self-attention networks have not yet introduced positional information, suggesting that these models have not yet fully exploited their potential. These considerations have motivated us to undertake this work. 

Despite the promising advancements in Quantum Machine Learning (QML), significant challenges persist in its practical implementation. One major hurdle is the inherent noise in current Noisy Intermediate-Scale Quantum (NISQ) devices \cite{preskill2018quantum}. While these quantum systems show potential for outperforming classical computers in specific tasks, they are significantly affected by quantum noise. The noise in quantum gates will limit the size of quantum circuits that can be executed reliably. Given these limitations, we primarily simulate quantum systems with noise models such as depolarizing, amplitude damping, and phase damping to evaluate performance under realistic quantum conditions.

In this study, we propose a novel Quantum Mixed-State Self-Attention Network (QMSAN), which is a hybrid quantum-classical model. This model integrates concepts from classical attention neural networks with the principle of mixed state from quantum computing, introducing an innovative attention mechanism. The core innovation of QMSAN lies in using quantum mixed states to represent queries and keys in the quantum attention mechanism, and directly calculating the similarity between these mixed states in the quantum domain through quantum swap tests. This approach not only improves the accuracy of similarity calculations but also fully leverages the advantages of quantum computing.
Our main contributions include:
\begin{itemize}
    \item We propose a novel method to calculate attention similarity between quantum mixed states keys and queries. This method first generates mixed states quantum queries and keys through partial trace operations of the quantum system, then directly calculates the similarity between these two mixed states using a swap test quantum circuit. This approach improves the accuracy and efficiency of similarity calculations.
    \item We introduce an innovative quantum positional information encoding method. This approach captures and encodes positional information within data through specifically designed fixed quantum gate operations, without the need for additional qubit resources, thereby enhancing its capability to capture long-range dependencies.
    \item Our QMSAN with positional encoding (QMSAN-P) consistently outperformed both the Quantum Self-Attention Neural Network (QSANN) and a Classical Self-Attention Neural Network (CSANN) \cite{li2024quantum} across the MC, RP, and Sentiment Labelled Sentences datasets. Additionally, the QMSAN-P model with positional encoding showed a 0.71\% to 1.08\% improvement in accuracy over the QMSAN without positional encoding (QMSAN-NP).
\end{itemize}

The rest of the paper is structured as follows: Section~\ref{section 2} summarizes the basic theory and methods. Section~\ref{section 3} explains QMSAN framework in detail and introduces the quantum circuits used in it. Section~\ref{section 4} describes our novel Quantum Position Encoding method. Section~\ref{section 5} outlines the Model Training process. Section~\ref{section 6} gives the numerical simulation setup and comparative results. Finally, Section~\ref{section 7} concludes the paper and discusses future directions.

\begin{table}[htbp]

\caption{Notations.}  
\label{tab:notations}  
\begin{tabular}{|l|l|}
\hline
Symbol & Description \\
\hline
$\ket{\psi}$ & Ket vector, representing a quantum pure state \\
\hline
$\bra{\psi}$ & Bra vector, Hermitian conjugate of $\ket{\psi}$ \\
\hline
$\braket{\phi|\psi}$ & Inner product, quantum state overlap \\
\hline
$\rho$ & Density matrix, describes mixed quantum states \\
\hline
$U$ & Unitary operator, represents quantum gates \\
\hline
$U^\dagger$ & Hermitian adjoint of $U$ (inverse operation) \\
\hline
$\otimes$ & Tensor product, used for composite quantum systems \\
\hline
$R_x(\theta), R_y(\theta), R_z(\theta)$ & Single-qubit rotation gates by angle $\theta$ about \\
 &$x$, $y$, $z$ axes of Bloch sphere \\
\hline
$\mathrm{tr}(\cdot)$ & Trace operation \\
\hline
$\mathrm{tr}_B(\cdot)$ & Partial trace operation over subsystem $B$ \\
\hline
$\mathcal{E}(\cdot)$ & Quantum channel, describes open system evolution \\
\hline
$M$ & Set of measurement operators \\
\hline
$I$ & Identity operator, represents no operation \\
\hline
$|\psi\rangle\langle\psi|$ & Projector onto state $|\psi\rangle$ \\
\hline
\end{tabular}
\end{table}

\section{{Preliminaries}}
\label{section 2}
Before delving into quantum self-attention networks, it's essential to understand some fundamental concepts of quantum mechanics.

\subsection{Quantum States}
Quantum systems can be described by pure states or mixed states.
Pure states are represented by $\ket{\psi}$ in Hilbert space. For a qubit:
\begin{equation}
\ket{\psi}=\alpha\ket{0}+\beta\ket{1},
\end{equation}
where $|\alpha|^2+|\beta|^2=1$.

Mixed states describe probabilistic mixtures of pure states, represented by density matrices $\rho$:
\begin{equation}
\rho=\sum_ip_i\ket{\psi_i}\bra{\psi_i},
\end{equation}
where $\sum_i p_i = 1$. 

Quantum states evolve through unitary transformations. For pure states, the evolution is described as:
\begin{equation}
    \ket{\psi^{\prime}}=U\ket{\psi},
\end{equation}
where $U$ is a unitary operator acting on the state vector \( |\psi\rangle \).

For mixed states, represented by the density matrix $\rho$, the evolution follows a similar rule:
\begin{equation}
    \rho'=U\rho U^{\dagger},
\end{equation}
where \( U^\dagger \) is the Hermitian conjugate of the unitary operator \( U \).

\subsection{Observables in Quantum Mechanics}
Observables are used to extract classical information from quantum systems. An observable $M$ is represented by:
\begin{equation}
    M = \Sigma_i \lambda_i P_i,
\end{equation}
where $\lambda_i$ are  eigenvalues and $P_i$ are projection operators.

The expectation value of M for a pure state $\ket{\phi}$ is:

\begin{equation}
    \langle M \rangle = \sum_i \lambda_i \bra{\phi}P_i\ket{\phi},
\end{equation}

For mixed states:
\begin{equation}
    \langle M\rangle=\mathrm{tr}(\rho M),
\end{equation}
where $\text{tr}(\cdot)$ represents the trace operation.  In this paper, we will use the observable $Z$ where
$Z=\left(+1\right)\ket{0}\bra{0}+\left(-1\right)\ket{1}\bra{1}$. In a system of $n$ qubits, the observable $n$ for the first qubit is mathematically expressed as $Z_{1}=Z\otimes I^{\otimes(n-1)}$.
\section{Quantum Mixed-State Attention Network Framework}
\label{section 3}
\subsection{Overview of QMSAN Architecture}
\begin{figure*}[h]
	\centering
	\includegraphics[width=1\textwidth]{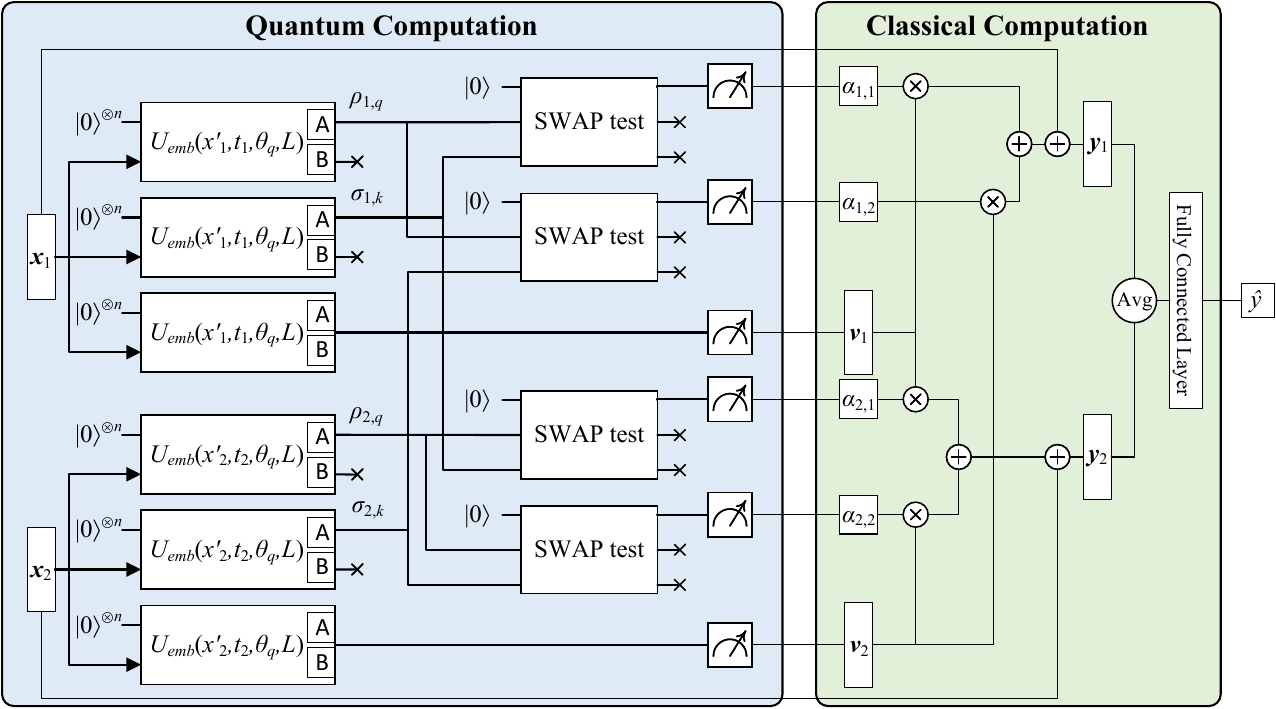}
	\caption{Quantum Mixed-State Self-Attention Network Framework.}
	\label{fig:overall}
\end{figure*}

The framework of QMSAN is illustrated in Fig.\,\ref{fig:overall}. At its core, QMSAN employs a three-stage quantum-classical hybrid process. First, classical input data is transformed into quantum information using trainable quantum embedding circuits, projecting the information into a high-dimensional Hilbert space. This quantum feature mapping generates three distinct quantum representations: mixed states for queries and keys ($\rho_{s,q}$ and $\sigma_{s,k}$), and pure states for values ($\ket{x_{s,v}}$). The model uses quantum operations to compute attention coefficients, with the swap test quantum circuit determining the similarity between mixed states, which is then converted into classical data. This classical similarity data is then integrated with measurements from the value states using the Pauli-$Z$ observable, and a fully connected neural network layer completes the final binary classification.

Recognizing the critical role of similarity computation between keys and queries in classical attention mechanisms, we leverage mixed quantum states for this calculation in our quantum attention design. By employing mixed states, we aim to leverage the potentially richer similarity computations they facilitate, compared to pure states, which may contribute to enhanced performance in quantum attention mechanisms.

In quantum self-attention networks, a straightforward and intuitive method for calculating the similarity between queries and keys is to use the inner product of their corresponding pure states. This technique known as 'quantum kernels' in quantum machine learning \cite{schuld2021supervised,li2024quantum}:
\begin{equation}
    \alpha_{s, j} = \left| \langle x_{s,q} | x_{j,k} \rangle \right|^2.
\end{equation}

However, in quantum systems with the same number of qubits, when dealing with mixed states, we can use a more expressive method for calculating similarity. For mixed states $\rho_{s,q}$ and $\sigma_{j,k}$, we can define their similarity as 
\begin{equation}
    \alpha_{s, j} = \operatorname{tr}(\rho_{s,q}\sigma_{j,k}),
\end{equation}
where $\operatorname{tr}(\cdot)$ denotes the trace operation. The trace-based similarity measure for mixed states offers a more comprehensive and nuanced comparison than the inner product method used for pure states. 

This approach captures a broader range of quantum state relationships. To illustrate this, we consider two arbitrary mixed states: $\rho_1 = \sum_i p_i |\psi_i\rangle\langle\psi_i|$, $\rho_2 = \sum_j q_j |\phi_j\rangle\langle\phi_j|$. The trace-based similarity measure between two mixed states can be expressed as $\text{tr}(\rho_1\rho_2)$:
\begin{equation}
\begin{split}
\text{tr}(\rho_1\rho_2) 
&= \text{tr}\left(\sum_i p_i |\psi_i\rangle\langle\psi_i| \sum_j q_j |\phi_j\rangle\langle\phi_j|\right) \\
&= \sum_i\sum_j p_i q_j \text{tr}(|\psi_i\rangle\langle\phi_j| \langle\psi_i|\phi_j\rangle) \\
&= \sum_i\sum_j p_i q_j \langle\phi_j|\psi_i\rangle \langle\psi_i|\phi_j\rangle \\
&= \sum_{i,j} p_i q_j |\langle\psi_i|\phi_j\rangle|^2.
\end{split}
\end{equation}

By allowing weighted combinations of multiple pure states overlaps, the trace-based similarity measure for mixed states offers a richer representation of quantum state relationships compared to the pure states inner product.  

\begin{figure*}[htbp]
    \centering
    \subfigure[Pure State]{
        \begin{minipage}[b]{0.35\textwidth}
            \includegraphics[width=\linewidth]{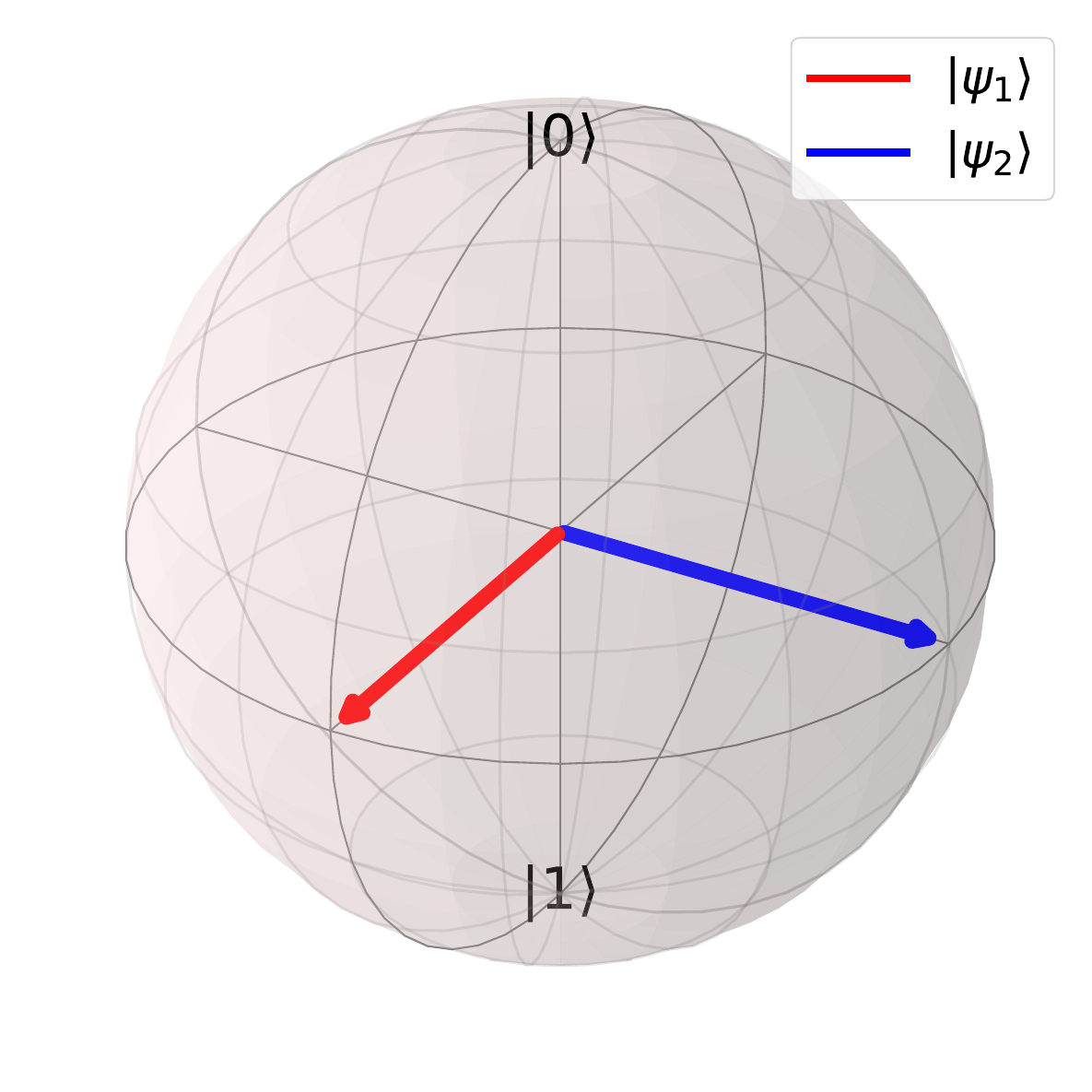}
        \end{minipage}
        \label{fig:bloch_sphere_pure}
    }
    \subfigure[Mixed State]{
        \begin{minipage}[b]{0.35\textwidth}
            \includegraphics[width=\linewidth]{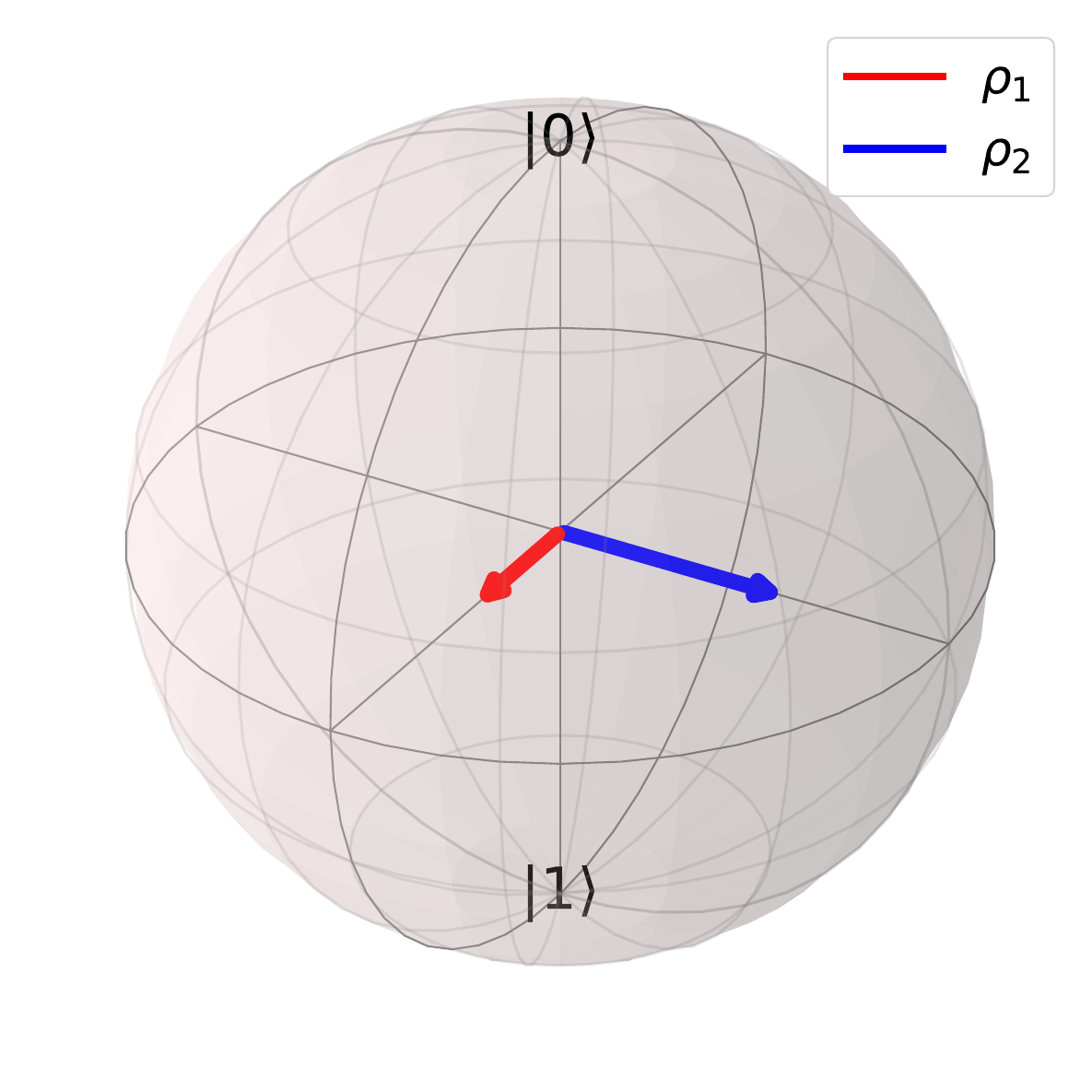}
        \end{minipage}
        \label{fig:bloch_sphere_mixed}
    }
    \caption{Comparison of Pure and Mixed States on Bloch Sphere.}
    \label{fig:bloch_spheres}
\end{figure*}

Moreover, visualizing quantum states on the Bloch sphere highlights the enhanced  expressivity of mixed state similarity measures compared to pure state measures, using a simple single-qubit example. As shown in Fig.\,\ref{fig:bloch_spheres}, pure states are confined to the sphere's surface, differing only in rotation. In contrast,In contrast, mixed states are represented by points within the sphere's volume, not restricted to its surface. This geometric representation suggests that mixed state similarity measures may capture a more diverse range of quantum state relationships compared to pure state inner products. The application of mixed states similarity measures in quantum attention neural networks may allow for the exploration of more diverse quantum state relationships.

\subsection{Quantum Mixed State Embedding}\label{AA}
Based on the preceding analysis, we propose a quantum self-attention network based on mixed states. There are two key issues we need to address. 

First is encoding classical data into quantum Hilbert space. We employ an iterative trainable quantum circuit architecture that alternates between data embedding and variational quantum circuits. This structure effectively encodes classical input data into quantum states, projecting it into a high-dimensional Hilbert space. Second is to produce mixed states. we leverage partial trace operations on pure states to produce mixed states, enabling a more comprehensive similarity calculation between queries and keys.

\subsubsection{Quantum Embedding Circuit}
Traditionally, quantum machine learning approaches have relied on fixed quantum feature maps followed by trainable variational circuits. This architecture, combining a data encoding circuit with a learnable Quantum Neural Network (QNN), has been widely adopted in various quantum machine learning models \cite{cong2019quantum, li2023quantum, schuld2020circuit, ostaszewski2021reinforcement, ostaszewski2021structure}. However, recent studies \cite{schuld2021effect, lloyd2020quantum, li2022concentration} suggest that this approach may not be optimal. This approach has limitations. Specifically, it necessitates meticulous design of the initial fixed quantum encoding circuit, as this component significantly influences the model's overall performance and generalization capabilities. The encoding circuit essentially determines the quantum feature space in which the data is represented, and its design can greatly impact the effectiveness of subsequent quantum operations.

\begin{figure*}[h]
	\centering
	\subfigure[Ring (R)]{
		\begin{minipage}[b]{0.45\textwidth}
			\includegraphics[height=0.6\textwidth]{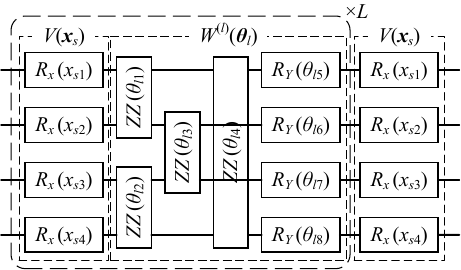} 
		\end{minipage}
		\label{fig:grid_4figs_1cap_4subcap_1}
	}
	\subfigure[Circuit-block (CB)]{
		\begin{minipage}[b]{0.45\textwidth}
			\includegraphics[height=0.6\textwidth]{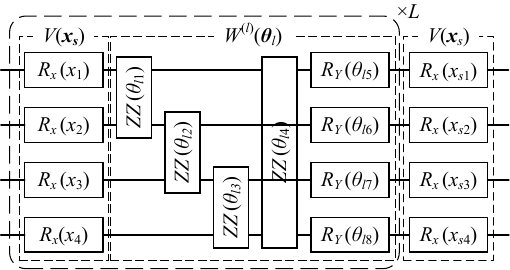}
		\end{minipage}
		\label{fig:grid_4figs_1cap_4subcap_2}
	}
	\\ 
	\subfigure[All-to-all (AA)]{
		\begin{minipage}[b]{0.5\textwidth}
			\includegraphics[height=0.5\textwidth]{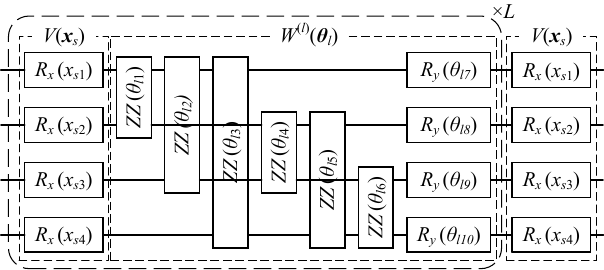} 
		\end{minipage}
		\label{fig:grid_4figs_1cap_4subcap_3}
	}
	\caption{Circuits considered for comparing two-qubit interaction configurations.}
	\label{fig:quantum embeddings}
\end{figure*}

In our quantum self-attention network, we adopt an iterative architecture for quantum embedding, as introduced by Ref.~\cite{lloyd2020quantum}. This design, inspired by ‘feature extractors’ in classical machine learning \cite{chen2016deep}, incorporates trainable elements throughout the encoding process. To evaluate its effectiveness, we compare three common circuit configurations for quantum embedding, as shown in Fig.\,\ref{fig:quantum embeddings}. Mirroring classical self-attention networks, our model employs three distinct quantum embeddings for query, key, and value components, sharing the same circuit configuration but with different trainable parameters. These embeddings transform classical input data $\boldsymbol{x}_s$ into quantum states $\ket{x_{s,q}}$, $\ket{x_{s,k}}$, and $\ket{x_{s,v}}$, where $1\leq s\leq S$, and $S$ denotes the number of input vectors in the data sample.

Specifically, these quantum embeddings use the same circuit configuration, implemented with different parameters $\boldsymbol{\theta}_q$, $\boldsymbol{\theta}_k$, and $\boldsymbol{\theta}_v$ for query, key, and value functions, respectively. The circuit employs single qubit and two qubit quantum gates. First, we use the single qubit gate $R_x(x_i)$ to encode the input data $\boldsymbol{x}=\left(x_{1}, \ldots, x_{N}\right)^{T}$ into the quantum circuit. Then, we use the $R_{zz}(\theta_1)=e^{-i \theta_{1} \sigma_{z} \otimes \sigma_{z}}$ gate to entangle the qubits and add $R_y(\theta_2)$. To enhance the expressivity of the quantum circuit, it can contain $L$ layers of such structure. Finally, we add the single qubit gate $R_x(x_i)$ again in the last layer to encode the data. Thus, the entire circuit can be represented as $U_{emb}(\boldsymbol{x},\boldsymbol{\theta})$, where $\boldsymbol{x}$ is the input data and $\boldsymbol{\theta}$ are the trainable parameters. Each layer consists of a data encoding circuit block $V(\boldsymbol{x})$ and a trainable circuit block $W(\boldsymbol{\theta}_l)$ controlled by the trainable parameters $\boldsymbol{\theta}_l$ of each layer.
\begin{equation}
U_{emb}(\boldsymbol{x},\boldsymbol{\theta},L)=V(\boldsymbol{x})\prod_{l=1}^L\left(W^{(l)}(\boldsymbol{\theta}_l)V(\boldsymbol{x})\right).
\end{equation}

Therefore, through three trainable quantum embeddings, we embed the input data $\boldsymbol{x}_s$ into three quantum states:
\begin{equation}
\begin{split}
\ket{x_{s,q}}=U_{emb}( \boldsymbol{x}_s,\boldsymbol{\theta}_q,L )\ket{0}^{\otimes n},\\
\ket{x_{s,k}}=U_{emb}( \boldsymbol{x}_s,\boldsymbol{\theta}_k,L )\ket{0}^{\otimes n},\\
\ket{x_{s,v}}=U_{emb}( \boldsymbol{x}_s,\boldsymbol{\theta}_v,L )\ket{0}^{\otimes n}.
\end{split}
\end{equation}

\subsubsection{Transformation to Mixed Quantum State}

For $\ket{x_{s,q}}$ and $\ket{x_{s,k}}$, they are obtained from the initial state $\ket{0}^{\otimes n}$ through the unitary transformation $U_{emb}(\boldsymbol{x},\boldsymbol{\theta})$, so $\ket{x_{s,q}}$ and $\ket{x_{s,k}}$ are both pure states. To obtain the mixed states, we extract information from the first $n/2$-qubit subsystem $A$ of the entire $n$-qubit quantum system by performing a partial trace operation on the quantum system and discarding the remaining n/2-qubit subsystem $B$. Specifically, this operation transforms the pure states $\ket{x_q}$ and $\ket{x_k}$ of the entire system into the mixed states $\rho_q$ and $\sigma_k$ of the corresponding subsystems, respectively:
\begin{equation}
	\begin{aligned}
		\rho_{s,q}=\operatorname{tr}_{B}(\ket{x_{s,q}}\bra{x_{s,q}}),\\
		\sigma_{s,k}=\operatorname{tr}_{B}(\ket{x_{s,k}}\bra{x_{s,k}}),
	\end{aligned}
\end{equation}
where $\operatorname{tr}_{B}(\cdot)$ is the partial trace over system $B$.

When measuring similarity between these mixed states, we can capture more nuanced relationships than those possible with pure state inner products. This approach allows us to quantify differences not just in the quantum states' orientations, but also in their degrees of mixture and entanglement. Consequently, the similarity measure between mixed states provides a more comprehensive comparison, potentially leading to more effective attention mechanisms in our quantum self-attention network.

\subsection{Quantum Self-Attention Mechanism}
\label{3.b}

Our quantum self-attention mechanism computes the similarity between queries and keys by leveraging quantum operations, specifically the swap test, before performing any measurement. Unlike traditional methods that measure quantum states of queries and keys separately to obtain classical data for similarity calculations \cite{li2024quantum}, our method performs quantum operations on the joint state of queries and keys before measurement. We aim to preserve more of the quantum information and correlations, potentially leading to more accurate and efficient similarity assessments in our quantum self-attention network.

Our approach to computing attention coefficients for mixed state queries and keys is inspired by the Hilbert-Schmidt distance \cite{coles2019strong,lloyd2020quantum}. Its definition is as follows:
\begin{equation}
D_{\mathrm{HS}}(\rho, \sigma)=\operatorname{tr}\left((\rho-\sigma)^{2}\right).
\end{equation}

Expanding the Hilbert-Schmidt distance equation yields three terms: $\operatorname{tr}(\rho\sigma)$, $\operatorname{tr}(\sigma^2)$, and $\operatorname{tr}(\rho^2)$ \cite{lloyd2020quantum}. The $\operatorname{tr}(\rho\sigma)$ term quantifies the overlap between two quantum ensembles in Hilbert space, with values ranging from 0 (orthogonal states) to 1 (identical pure states). The 'purity' terms $\operatorname{tr}(\rho^2)$ and $\operatorname{tr}(\sigma^2)$ measure intra-cluster overlap. For our quantum self-attention network, we focus solely on $\operatorname{tr}(\rho\sigma)$ to compute query-key similarities, omitting the purity terms as they are less relevant to inter-state comparisons.

We define quantum self-attention coefficient between the $s$-th and $j$-th mixed states, computed from the corresponding query and key parts:
\begin{equation}
    \label{equ.13}
    \alpha_{s,j}=\operatorname{tr}(\rho_{s,q}\sigma_{j,k}).
\end{equation}

\begin{figure}[h]
\centerline{\includegraphics{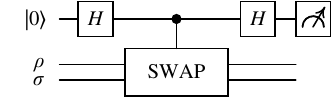}}
\caption{Quantum circuit implementing the SWAP test.}
\label{fig.SWAPtest}
\end{figure}

The above equation can be easily implemented by the SWAP test quantum circuit  \cite{garcia2013swap,kobayashi2003quantum}, as shown in Fig.\,\ref{fig.SWAPtest}. Following this, we outline the basic principles behind the circuit's implementation of Equation\ref{equ.13}.

Suppose we have a pair of mixed states $\rho$ and $\sigma$ of $n$ qubits, with $\rho=\sum_ip_i\ket{e_i}\bra{e_i}$ and $\sigma=\sum_iq_i\ket{f_i}\bra{f_i}$ decomposed using their respective orthogonal bases ${\ket{e_i}}$ and ${\ket{f_i}}$. If we perform a measurement on the auxiliary qubit and obtain the result $\ket{0}$, the SWAP test passes, otherwise it fails. The probability of the mixed states $\rho\otimes\sigma$ passing the SWAP test \cite{niu2022entangling} is :
\begin{equation}
\begin{aligned}
p(\ket{0})&=\sum_i\sum_jp_iq_j\left(\frac12+\frac{|\langle e_i|f_j\rangle|^2}2\right)=\frac12+\frac12\mathrm{tr}(\rho\sigma). 
\end{aligned}
\end{equation}

Therefore, we use the SWAP test quantum circuit to implement the calculation of quantum self-attention coefficients between queries and keys. This can effectively estimate the closeness of two mixed states. If the two mixed states are identical, $\rho = \sigma$, the test always passes with $p=1$. When the states are different, the finite probability $p$ of passing the test depends on the similarity $\operatorname{tr}{(\rho\sigma)}$ between the two states; the closer they are, the greater the probability of passing the test.

Additionally, the SWAP test can also efficiently compute the overlap between two pure states. For pure states $|\psi\rangle$ and $|\phi\rangle$, the probability of measuring $|0\rangle$ in the auxiliary qubit is:
\begin{equation}
    p(|0\rangle) = \frac{1}{2} + \frac{1}{2}|\langle\psi|\phi\rangle|^2
\end{equation}

This simpler case directly measures the squared inner product between the two quantum states. The relationship between pure state overlap and mixed state trace becomes clear when we consider pure states as special cases of mixed states, where $\rho = |\psi\rangle\langle\psi|$ and $\sigma = |\phi\rangle\langle\phi|$. This duality highlights the SWAP test's versatility in quantum similarity measurements, from pure state overlaps to more general mixed state comparisons.

The output solution process is described in matrix form, and coefficients matrix can be represented as
\begin{equation}
    \boldsymbol{C}=\begin{bmatrix}\tilde{\alpha}_{1,1}&\tilde{\alpha}_{1,2}&\cdots&\tilde{\alpha}_{1,S}\\
    \tilde{\alpha}_{2,1}&\tilde{\alpha}_{2,2}&\cdots&\tilde{\alpha}_{2,S}\\
    \vdots&\vdots&\ddots&\vdots\\
    \tilde{\alpha}_{S,1}&\tilde{\alpha}_{S,2}&\cdots&\tilde{\alpha}_{S,S}
\end{bmatrix},
\end{equation}
where $\tilde{\alpha}_{s,j}$ represents the normalized quantum self-attention coefficients:
\begin{equation}
\tilde{\alpha}_{s,j}=\frac{\alpha_{s,j}}{\sum_{m=1}^S\alpha_{s,m}},
\end{equation}

For the value part, we use an $n$-dimensional vector to represent it, with the observable $Z$ measured for each qubit, resulting in a vector with the same dimension as the number of qubits.
\begin{equation}
\boldsymbol{v}_s=\begin{bmatrix}\langle Z_{s,1}\rangle&\langle Z_{s,2}\rangle&\cdots&\langle Z_{s,n}\rangle\end{bmatrix}^\top,
\end{equation}

Finally, we adopt the structure of a residual network to design the output $y_s$, aiming to prevent network degeneration:
\begin{equation}
\boldsymbol{y}_s=\boldsymbol{x}_s+\sum_{j=1}^{S} {\boldsymbol{C}_{s,j}\cdot\boldsymbol{v}_j}.
\end{equation}

\section{Quantum Position Encoding}
\label{section 4}
While our QMSAN, similar to classical self-attention networks, excels at modeling token relationships and capturing contextual representations, it inherently lacks the capability to distinguish the sequential order of input tokens \cite{wu2021rethinking, su2024roformer, yun2019transformers}. This limitation is intrinsic to the attention mechanism. To overcome this and enable the model to leverage token positions, we need to explicitly incorporate positional information into the input sequence.
\begin{figure}[h]
    \centering
    \includegraphics[width=0.6\textwidth]{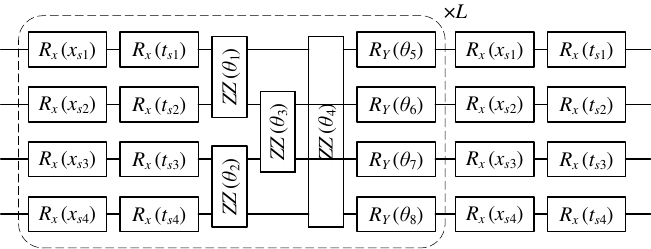}
    \caption{Quantum Positional Encoding in Ring Configuration.}
    \label{fig:position}
\end{figure}

Previous approaches to quantum positional encoding, such as the method proposed by Ref.~\cite{Chen2024Quantum}, require additional qubits to encode positional information. This approach, while effective, increases the quantum resource requirements. In contrast, our approach leverages the existing qubit structure more efficiently.
As illustrated in Fig.\,\ref{fig:position}, we introduce a novel quantum circuit design that incorporates positional information without auxiliary qubits. Our approach encodes positional data by adding single-qubit $R_x(\cdot)$ rotation gates.

To integrate positional information into our quantum circuit, we adapt the sinusoidal positional encoding introduced by Ref.~\cite{vaswani2017attention} for the classical Transformer model. In the classical approach, the positional encoding for even and odd dimensions is defined as:
\begin{equation}
\begin{aligned}
PE_{s,2i}&=sin(s/10000^{2i/d_{\text{model}}}),\\
PE_{s,2i+1}&=cos(s/10000^{2i/d_{\text{model}}}),
\end{aligned}
\end{equation}
where $s$ represents the position, $i$ is encoding vector dimension index, and $d_{\text{model}}$ the embedding dimension.
To adapt this encoding for our quantum circuit, we perform distinct scaling operations on the positional encoding and input data:
\begin{equation}
\boldsymbol{t}_s=\frac{\boldsymbol{PE}_{s}-PE_{min}}{PE_{max}-PE_{min}}\times2\pi,
\end{equation}
\begin{equation}
\boldsymbol{x}'_s=\frac{\boldsymbol{x}_s-x_{min}}{x_{max}-x_{min}}\times\pi,
\end{equation}
where $x_{min}$ and $x_{max}$ are the global minimum and maximum values across all input vectors, while $PE_{min}$ and $PE_{max}$ are the extrema of the positional encoding data.

This differentiated scaling approach is necessitated by the unique properties of quantum rotation gates, specifically the $R_x(\theta)$ gate, which rotates the quantum state around the $x$-axis by an angle $\theta$. The $R_x$ gate has a natural periodicity of $2\pi$, which aligns well with the inherent periodicity of the sinusoidal positional encoding. Consequently, we scale the positional encoding to the full $[0, 2\pi]$ range to exploit this alignment.

In contrast, the input data $\boldsymbol{x}_s$ is non-periodic. Scaling it to the full $[0, 2\pi]$ range could introduce unintended symmetries in the quantum representation, as evidenced by the relationship $R_x(2\pi - x'_s) = -Z R_x(x'_s) Z$. This symmetry, while natural for periodic positional encoding, doesn't reflect the nature of the original input data. Therefore, to avoid introducing these unintended symmetries and to preserve the non-periodic characteristics of the original input data, we restrict the scaling of $\boldsymbol{x}_s$ to the $[0, \pi]$ range in its quantum representation.

These tailored scaling strategies enable us to map both positional encoding and input data to quantum states via rotation gates $R_x(\cdot)$, preserving their distinct periodic and non-periodic characteristics. This approach maintains the integrity of the original data structure in the quantum representation, potentially benefiting subsequent quantum machine learning tasks.

\section{Model Training}
\label{section 5}
We train our model on the dataset $\mathcal{D}=\{(\boldsymbol{x}_{m;1},\boldsymbol{x}_{m;2},\ldots,\boldsymbol{x}_{m;S_m}),y_m\}_{m=1}^{N_s}$ by minimizing the loss function, where $N_s$ represents the total number of samples, and the label $\Bar{y}_m\in{\{0,1\}}$ for each sample indicates its category. 

For each sample, $S_m$ denotes the number of words it contains, and each input data $\boldsymbol{x}_{m,s}$ is an $n$-dimensional vector.

The feature vector for each sample is obtained by summing and averaging the outputs $y_{m,s}$, where $1\leq m\leq N_{s}$ and $1\leq s\leq S_{m}$:
\begin{equation}
\boldsymbol{y}_{m}=\frac{1}{S_m}\sum_{s=1}^{S_{m}}\boldsymbol{y}_{m,s}.
\end{equation}

The output $\boldsymbol{y}_{m}$ of each sample is fed into a fully connected layer to produce the binary prediction value $\hat{y}_m$ for each sample.
\begin{equation}
\hat{y}_m:=\text{Sigmoid}(\boldsymbol{w}^\top\cdot \boldsymbol{y}_{m}+{b}),
\end{equation}
where $\boldsymbol{w}$ and $b$ represent the weight and bias of the fully connected layer, respectively, and $\text{Sigmoid}$ denotes the sigmoid activation function.

For classification tasks, there are many loss functions to choose from, such as cross-entropy loss and mean squared error (L2 loss). In the current work, we use the simple and effective mean squared error as the loss function:
\begin{equation}
\mathcal{L}\left(\boldsymbol{\Theta},\boldsymbol{w},{b};\mathcal{D}\right)=\frac1{2N_s}\sum_{m=1}^{N_s}\left(\hat{y}_m-\Bar{y}_m\right)^2,
\end{equation}
where $\boldsymbol{\Theta}$ represents all trainable parameters in the quantum circuits.

\begin{algorithm}
\footnotesize

    \renewcommand{\algorithmicrequire}{\textbf{Input:}}
    \renewcommand{\algorithmicensure}{\textbf{Output:}}
    \caption{QMSAN training algorithm.}
    \begin{algorithmic}[1]
        \REQUIRE{Batch sizes $\textbf{BS}$. Number of words per sample $S_{m}$. Learning rate $\eta$. Number of quantum embedding Layers $L$. Number of qubits $n$. The scaled position encodings $t_{s}$. The scaled training data set $\mathcal{D}=(\boldsymbol{x'}_{m;1},\boldsymbol{x'}_{m;2},\ldots,\boldsymbol{x'}_{m;S_m}),y_m\}_{m=1}^{N_s}$.\\
        $\boldsymbol{\Theta}\sim\mathcal{N}(0,0.01)$, $\boldsymbol{w}\sim\mathcal{N}(0,0.01)$, $b\leftarrow 0$}
        \REPEAT 
        \FOR{$m$ from 1 to $\textbf{BS}$} 
                \FOR{$s$ from 1 to $S_m$}
                        \STATE $\rho_{m,s,q} \gets \operatorname{tr}_{B} ( U_{\text{emb}}(\boldsymbol{x}'_{m,s}, \boldsymbol{t}_s, \boldsymbol{\theta}_q, L) \ket{0}^{\otimes n} \bra{0}^{\otimes n} U^\dagger_{\text{emb}}(\boldsymbol{x}'_{m,s}, \boldsymbol{t}_s, \boldsymbol{\theta}_q, L) )$
                        \STATE $\rho_{m,s,k} \gets \operatorname{tr}_{B} ( U_{\text{emb}}(\boldsymbol{x}'_{m,s}, \boldsymbol{t}_s, \boldsymbol{\theta}_k, L) \ket{0}^{\otimes n} \bra{0}^{\otimes n} U^\dagger_{\text{emb}}(\boldsymbol{x}'_{m,s}, \boldsymbol{t}_s, \boldsymbol{\theta}_q, L) )$
                        \STATE $\ket{x_{m,s,v}} \gets U_{emb}(\boldsymbol{x}'_{m,s},\boldsymbol{t}_s,\boldsymbol{\theta}_v,L)\ket{0}^{\otimes n}$
                        \STATE $v_{m,s} \gets \begin{bmatrix}\langle Z_{m,s,1}\rangle&\langle Z_{m,s,2}\rangle&\cdots&\langle Z_{m,s,n}\rangle\end{bmatrix}^\top$
                \ENDFOR
                \FOR{$s$ from 1 to $S_m$}
                    \FOR{$j$ from 1 to $S_m$}
                        \STATE $\alpha_{m,s,j} \gets \text{tr}(\rho_{m,s,j} \sigma_{m,j,k})$  
                    \ENDFOR
                    \STATE $\text{SUM}_{m,s} \gets \sum_{j=1}^{S_m} \alpha_{m,s,j}$
                    \FOR{$j$ from 1 to $S_m$}
                        \STATE $\alpha_{m,s,j} \gets \alpha_{m,s,j} / \text{SUM}_{m,s}$
                    \ENDFOR
                \ENDFOR

                \STATE $y_m \gets \sum_{s=1}^{S_m} \sum_{j=1}^{S_m} \alpha_{m,s,j} \cdot v_{m,j}$  

                \STATE $\hat{y}_m:=\text{Sigmoid}(\boldsymbol{w}^\top\cdot \boldsymbol{y}_{m}+{b})$
        \ENDFOR
        \STATE $\mathcal{L}\gets\frac{1}{2N_s}\sum_{m=1}^{N_s}\left(\hat{y}_m-\Bar{y}_m\right)^2 $
        \STATE $\boldsymbol{\Theta}\gets\boldsymbol{\Theta}-\eta\nabla_{\boldsymbol{\boldsymbol{\Theta}}}\mathcal{L}$; $\boldsymbol{w}\gets\boldsymbol{w}-\eta\nabla_{\boldsymbol{w}}\mathcal{L}$; ${b}\gets {b}-\eta\nabla_{{b}}\mathcal{L}$
        \UNTIL{$\mathcal{L}$ converges or the number of iterations reaches the maximum}
        \ENSURE{Optimal parameters $\boldsymbol{\Theta}^{*}$,$\boldsymbol{w}^{*}$,${b}^{*}$ }
    \end{algorithmic}
\end{algorithm}
\section{Numerical Experiments}
\label{section 6}
To evaluate our model's capabilities, we utilized two types of publicly available datasets: simple sentence datasets (MC and RP \cite{lorenz2023qnlp}) for testing basic language understanding, and sentiment analysis datasets \cite{misc_sentiment_labelled_sentences_331} (Yelp, IMDb, and Amazon reviews) for evaluating more complex natural language processing tasks. Our experiments included three main components. Initially, we evaluated models without positional encoding across all datasets to establish a baseline. Subsequently, we focused on the impact of adding positional encoding in sentiment analysis tasks, allowing us to evaluate its effectiveness in more complex natural language processing contexts. Finally, to simulate the challenges posed by noisy intermediate-scale quantum (NISQ) devices, we evaluated the model's robustness under various quantum noise types, including depolarizing channel, phase damping, and amplitude damping.

\subsection{Datasets and Experimental Setup}\label{SCM}
\subsubsection{Datasets}

Our study evaluates the Quantum Mixed-State Self-Attention Network (QMSAN) using three datasets of increasing complexity:
\begin{itemize}
    \item Meaning Classification (MC): 130 short sentences(70 train + 30 development + 30 test) split between food and IT topics, using a 17-word vocabulary.
    \item RELPRON (RP): 105 four-word noun phrases(74 train + 31 test) with relative clauses and a 115-word vocabulary.
    \item Sentiment Labelled Sentences dataset: 1000 reviews each from Amazon, IMDb, and Yelp, with balanced positive and negative sentiments. Vocabulary sizes: Amazon 1906, IMDb 3173, Yelp 2081. Each dataset is randomly split into 80\% training and 20\% test sets.
\end{itemize}

\subsubsection{Model Configuration}
In our work, we use the Tensorcircuit framework \cite{zhang2023tensorcircuit} for simulating quantum circuits and the Tensorflow \cite{abadi2016tensorflow} framework for parameter optimization, with the optimizer Adam  \cite{kingma2014adam}. Our experimental setup uses a batch size of 64, with training continuing until convergence. To ensure robustness, we conduct 15 runs with different initializations for MC and RP datasets, while employing 3 runs of 5-fold cross-validation for the Sentiment Labelled Sentences dataset. Parameters are initialized from a normal distribution ($\mu$=0, $\sigma$=0.1), and Pauli-$Z$ observables are used for quantum measurements. For fair comparison, we use the same qubit configurations as in Ref.~\cite{li2024quantum}: 2 qubits for MC and 4 for other tasks. Detailed hyperparameter settings are presented in Table \,\ref{table:experiment-config}.

\begin{table}[htbp]

\caption{Comprehensive Experimental Configuration}
\label{table:experiment-config}
\centering
\resizebox{\textwidth}{!}{%
\begin{tabular}{lccccccccc}
\toprule
\multirow{2}{*}{Dataset} & \multirow{2}{*}{Qubits ($n$)} & \multirow{2}{*}{Layers ($L$)} & \multirow{2}{*}{Sequence Length} & \multicolumn{3}{c}{QMSAN-NP Learning Rate} & \multicolumn{3}{c}{QMSAN-P Learning Rate} \\
\cmidrule(lr){5-7} \cmidrule(lr){8-10}
 &  &  &  & R & CB & AA & R & CB & AA \\
\midrule
MC    & 2 & 1 & 4 & 0.005 & 0.006 & 0.009 & / & / & / \\
RP    & 4 & 2 & 4 & 0.002 & 0.050 & 0.010 & / & / & / \\
IMDb  & 4 & 1 & 45 & 0.008 & 0.008 & 0.010 & 0.008 & 0.008 & 0.008 \\
Yelp  & 4 & 1 & 32 & 0.007 & 0.007 & 0.030 & 0.030 & 0.030 & 0.010 \\
Amazon & 4 & 1 & 30 & 0.080 & 0.009 & 0.090 & 0.020 & 0.010 & 0.020 \\
\bottomrule
\multicolumn{10}{l}{\textit{Note:}R: Ring, CB: Circuit-block, AA: All-to-all. -NP: without positional encoding, -P: with positional encoding.}\\
\multicolumn{10}{l}{Convergence criterion: gradient falls below $10^{-4}$ for 10 consecutive iterations. Batch Size: 64.}\\
\multicolumn{10}{l}{Adam optimizer used for all models. Parameters initialized from $\mathcal{N}(0, 0.1)$.} \\
\multicolumn{10}{l}{Hardware: Intel Core i9-12900H CPU, 16GB RAM.}\\
\end{tabular}
}
\end{table}

To explore circuit configurations effects, we implemented three entanglement schemes in quantum embeddings, as shown in Fig.\,\ref{fig:quantum embeddings}:

\begin{itemize}
    \item Ring (R) configuration: It derived from the Nearest-Neighbor(NN) configuration \cite{sim2019expressibility}. The two-qubit gates in the Ring configuration connect adjacent qubits in a circular loop, linking each qubit to its neighbors and finally connecting the first and last qubits to close the loop.
    \item Circuit-block (CB) configuration \cite{sim2019expressibility}: The two-qubit gates in the Circuit-block configuration connect adjacent qubits sequentially in each layer, and finally connect the first and last qubits in each layer to complete the circular loop.
    \item All-to-all (AA) configuration \cite{benedetti2019generative}: The two-qubit gates in the AA configuration form a fully connected graph, where each qubit is linked to every other qubit, allowing for maximum entanglement across the entire qubits.
\end{itemize}

These configurations form the basis of our model variants, denoted as \text{QMSAN-\{R$\vert$CB$\vert$AA\}-\{NP$\vert$P\}}, where NP and P indicate models without and with quantum positional encoding, respectively.

Experimental evaluation showed the All-to-all (AA) configuration's enhanced performance in both expressivity and entangling capability, while Circuit-block (CB) and Ring configurations show similar performance in both expressivity and entangling capability. For a comprehensive analysis, refer to \ref{app1}.

To compare the effectiveness of mixed states and pure states computations in quantum attention mechanisms, we introduce QPSAN, a variant of QMSAN-R-NP. This model is identical to QMSAN-R-NP in all aspects except one key difference: QPSAN uses pure states inner products for calculating similarity between query and key in the attention mechanism.

\subsubsection{Baseline Model}
Our study employs three baseline models for natural language processing tasks: two classical models, the DIStributional COmpositional CATegorical (DisCoCat) \cite{lorenz2023qnlp} and the Classical Self-Attention Neural Network (CSANN) \cite{li2024quantum}, as well as one quantum model, the Quantum Self-Attention Neural Network (QSANN) \cite{li2024quantum}.
\begin{itemize}
    \item DisCoCat: The DisCoCat model is a syntax-sensitive approach to NLP that represents words as tensors based on pregroup grammar types. It integrates distributional semantics with compositional structure, encoding sentences as string diagrams interpretable as vector space operations.
    \item CSANN: The CSANN is a classical model using a single self-attention layer and a fully-connected layer for binary classification, processing 16-dimensional inputs. It averages word embeddings to represent sentences before applying self-attention and classification.
    \item QSANN: The QSANN encodes classical data into high-dimensional quantum states using circuits structurally identical to the applied parameterized quantum circuits(PQCs), applies these PQCs following the self-attention layout, and uses Gaussian projected quantum self-attention (GPQSA) to generate outputs. The model concludes with a fully-connected layer for binary classification tasks.
\end{itemize}

\subsection{Basic Model Performance (Non-Positional)}
\subsubsection{Comparison with Classical Models}
Our QMSAN-NP models demonstrate significant improvements over classical approaches like DisCoCat and CSANN across various datasets. On the MC dataset, QMSAN-NP models achieve perfect classification accuracy, surpassing DisCoCat's 79.80\% accuracy. For the more complex RP dataset, QMSAN-AA-NP reaches 75.63\% accuracy, outperforming DisCoCat's 72.30\%. On sentiment analysis tasks, QMSAN-NP models consistently outperform CSANN, with improvements of up to 4.45\% on the IMDb dataset.
\begin{table*}[ht]
\caption{Test Accuracy Comparison on MC and RP Tasks.}
\label{table:TA MR}
\centering
\resizebox{\textwidth}{!}{%
\begin{tabular}{ccccccc}
\toprule
\multirow{2}*{Method} & \multicolumn{3}{c}{MC} & \multicolumn{3}{c}{RP}\\
\cmidrule(lr){2-4} \cmidrule(lr){5-7}
 & \#Paras & TrainAcc(\%) & TestAcc(\%) & \#Paras & TrainAcc(\%) & TestAcc(\%) \\
\midrule
DisCoCat \cite{lorenz2023qnlp} & $40$ & $83.10$ & $79.80$ & $168$ & $90.60$ & $72.30$ \\
QSANN \cite{li2024quantum} & 25 & $100.00$ & $100.00$ & $109$ & $95.35$ & $67.74$ \\
QPSAN & $15$ & $100.00$ & $100.00$ & $53$ & $96.10\pm0.7$ & $71.97\pm1.01$  \\
QMSAN-R-NP & $15$ & $100.00$ & $100.00$ & $53$ & $96.40\pm0.64$ & $74.91\pm2.03$  \\
QMSAN-CB-NP & $15$ & $100.00$ & $100.00$ & $53$ & $96.55\pm0.67$ & $74.55\pm1.83$  \\
QMSAN-AA-NP & $18$ & \pmb{$100.00$} & \pmb{$100.00$} & $137$ & \pmb{$96.85\pm0.64$} & \pmb{$75.63\pm1.6$}  \\
\bottomrule
\end{tabular}
}
\end{table*}

\begin{table*}[ht!]
\centering
\caption{Test Accuracy Comparison on Yelp, IMDb, and Amazon Datasets.}
\label{table:TA YIA}
\resizebox{\textwidth}{!}{%
\begin{tabular}{cccccccccc}
\toprule
\multirow{2}*{Method} & \multicolumn{3}{c}{Yelp} & \multicolumn{3}{c}{IMDb} & \multicolumn{3}{c}{Amazon}\\
\cmidrule(lr){2-4} \cmidrule(lr){5-7} \cmidrule(lr){8-10}
 & \#Paras & TrainAcc(\%) & TestAcc(\%) & \#Paras & TrainAcc(\%) & TestAcc(\%) & \#Paras & TrainAcc(\%) & TestAcc(\%) \\
\midrule
CSANN \cite{li2024quantum} & 785 &/& $83.11\pm0.89$ & 785 &/ & $79.67\pm0.83$ & 785 &/ & $83.22\pm1.28$ \\
QSANN \cite{li2024quantum} & 49 &/& $84.79\pm1.29$ & 49 &/& $80.28\pm1.78$ & 61 &/& $84.25\pm1.75$ \\
QPSAN          & 29 &$99.78\pm0.24$& $84.02\pm2.42$ & 29 &$99.65\pm0.34$& $83.82\pm1.20$ & 29 &$99.80\pm0.11$& $86.54\pm2.29$ \\
QMSAN-R-NP & 29 &$99.53\pm0.22$& $84.14\pm2.27$   & 29    &$99.48\pm0.37$& $84.12\pm2.31$     & 29 &$99.80\pm0.10$& $86.72\pm2.38$ \\
QMSAN-CB-NP & 29 &$99.58\pm0.23$& $84.40\pm1.98$  & 29   &$99.45\pm0.24$& $83.74\pm2.01$    & 29 &$99.83\pm0.17$& $86.61\pm1.71$ \\
QMSAN-AA-NP & 71 &$99.65\pm0.18$& $84.73\pm2.34$  & 71   &$99.50\pm0.40$& $83.76\pm3.04$    & 71 &$99.75\pm0.18$& $86.56\pm1.90$ \\

QMSAN-R-P & 29 &$99.45\pm0.32$& $84.85\pm1.33$           & 29   &$99.18\pm0.41$& $84.77\pm3.12$         & 29 &$99.87\pm0.94$& $87.41\pm1.16$ \\
QMSAN-CB-P & 29 &$99.80\pm0.20$& $84.82\pm1.21$           & 29   &$99.18\pm0.41$& \pmb{$84.82\pm2.96$}   & 29 &$99.90\pm0.93$& $87.43\pm1.16$ \\
QMSAN-AA-P & 71 &$99.55\pm0.26$& \pmb{$84.96\pm3.34$}     & 71   &$99.33\pm0.36$& $84.29\pm2.32$         & 71 &$99.91\pm0.50$& \pmb{$87.48\pm1.02$} \\
\bottomrule
\end{tabular}
}
\end{table*}
The QMSAN-NP quantum attention model demonstrates improved performance across various datasets compared to the classical attention model CSANN. A key factor in this performance is that quantum systems can efficiently map classical data to high-dimensional Hilbert spaces. We use angle encoding to map $n$-dimensional classical data into $n$ qubits, where the $n$ qubits represent a space of dimension $2^n$. This exponential increase in dimensionality enables the model to represent and process highly complex data patterns, potentially providing advantages in specific NLP tasks compared to classical neural networks. By accessing a much larger feature space, the quantum encoding approach allows for more detailed and potentially more powerful representations of the input data \cite{liu2021rigorous,lloyd2021quantum}.

The quantum attention model outperforms the classical DisCoCat model on the MC task, primarily because the task involves a smaller vocabulary, with the training and test sets exhibiting similar word choices and sentence structures. This consistency allows the attention mechanism to more effectively capture key words and patterns within sentences. However, on the RP dataset, both models exhibit similar performance levels. This is likely due to the fact that the RP task involves a larger vocabulary and more complex grammatical structures, while the training data remains limited. These factors may require more extensive data to fully leverage the advantages of an attention-based model. However, our QMSAN model enhances the encoding of quantum attention by utilizing a trainable quantum embedding and mixed states similarity calculations. By improving these details, our model captures finer nuances in the data, leading to slightly better performance on the RP dataset compared to the classical DisCoCat model.

\subsubsection{Comparison with Quantum Models}
When compared to QSANN, QMSAN-NP demonstrates improved performance across various datasets. On the RP dataset, QMSAN-AA-NP achieves 75.63\% accuracy, outperforming the 67.74\% accuracy of QSANN, as shown in Table \ref{table:TA MR}. For sentiment analysis tasks, QMSAN-NP models show improvements of up to 3.84\% on the IMDb dataset compared to QSANN, as shown in Table \ref{table:TA YIA}.

QMSAN-NP models show improvements over QSANN's method, which requires measuring the quantum states of queries and keys to convert them into classical data, and then calculating similarities using this classical information. In quantum mechanics, the act of measurement causes wave function collapse, irreversibly altering the quantum state. It can lead to loss of quantum information. By performing calculations directly within the quantum domain, QMSAN-NP maintains quantum state information, potentially capturing more subtle quantum correlations. QMSAN-NP can more effectively exploit the high-dimensional Hilbert space of the quantum system.

Compared to QPSAN, which differs from QMSAN-R-NP only in its use of pure states inner products for query-key similarity calculations, QMSAN-NP shows improved performance across various tasks.  As shown in Table \ref{table:TA MR}, QMSAN-AA-NP achieves 75.63\% accuracy on the RP dataset, outperforming QPSAN's highest accuracy of 73.48\%. Furthermore, as illustrated in Table \ref{table:TA YIA}, QMSAN-R-NP models consistently outperform QPSAN across all sentiment analysis datasets.

The enhanced performance of QMSAN-NP can be attributed to its use of mixed states similarity calculations rather than pure states inner products. The trace operation used in mixed states similarity calculations ($\operatorname{tr}(\rho\sigma)$) provides a more comprehensive measure of quantum state overlap compared to the inner product of pure states. This approach allows for a more nuanced assessment of quantum state relationships within the same Hilbert space dimension, potentially leading to more accurate attention mechanisms in quantum machine learning tasks.

\subsection{Experiments with Positional Models}

Across all datasets, QMSAN models with positional encoding (QMSAN-P) consistently outperform their counterparts without positional encoding (QMSAN-NP). As shown in Table \ref{table:TA YIA}, we observed accuracy improvements of 0.71\% on Yelp, 1.08\% on IMDb, and 0.92\% on Amazon datasets.

The improved performance of QMSAN-P can be attributed to our novel quantum positional encoding method. This approach leverages the same quantum rotation gates used for encoding classical inputs to encode fixed positional information. 

Exploiting the $2\pi$ periodicity of the $R_x(\theta)$ single-qubit quantum gate, we scale positional information to the full [0, $2\pi$] range. This allows us to fully utilize the natural cycle of quantum rotations for encoding positions. In contrast, we scale input data to the [0, $\pi$] range. This distinction is crucial as the input data lacks the intrinsic periodicity of positional information. Scaling input data to the full [0, $2\pi$] range could introduce unintended symmetries in the quantum representation, which don't reflect the nature of the original input data. By restricting input data scaling to [0, $\pi$], we preserve its non-periodic characteristics in the quantum representation.

\begin{figure*}[h]
    \centering
    \includegraphics[width=0.75\textwidth]{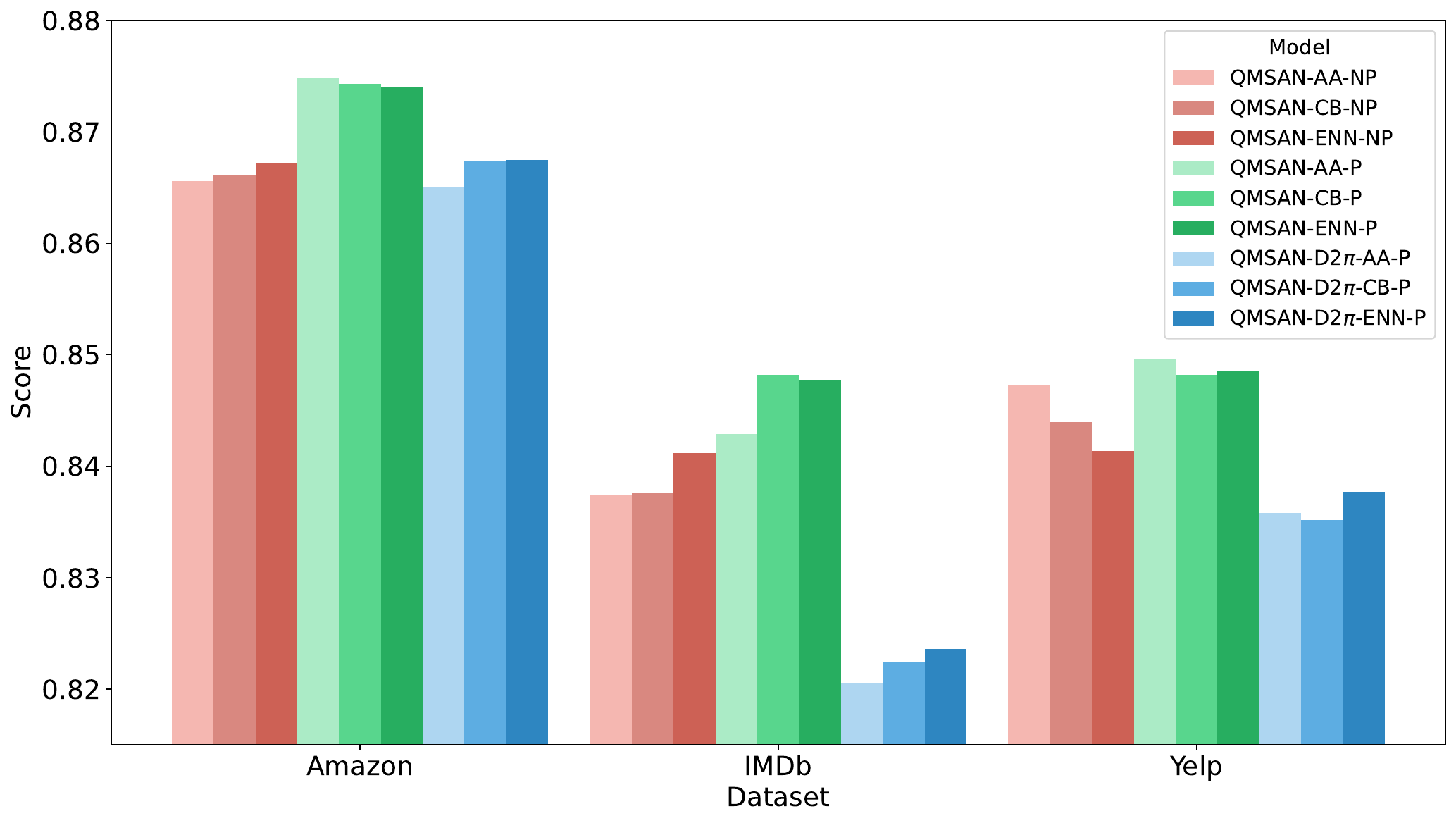}
    \caption{Test accuracy of different forms of data scaling methods.}
    \label{fig:Test accuracy}
\end{figure*}

Fig.\,\ref{fig:Test accuracy} illustrates that encoding both input data and positional information to the [0, $2\pi$] range yields lower performance across all datasets compared to our approach of scaling input data to [0, $\pi$] and positional information to [0, $2\pi$]. These results corroborate our theoretical analysis, demonstrating the importance of preserving the non-periodic nature of input data in quantum representations. By differentiating the scaling ranges for input and positional data, we effectively leverage the periodicity of quantum gates for positional encoding while maintaining the intrinsic characteristics of the input features.

\begin{figure}[h]
	\centering
	\includegraphics[width=1\textwidth]{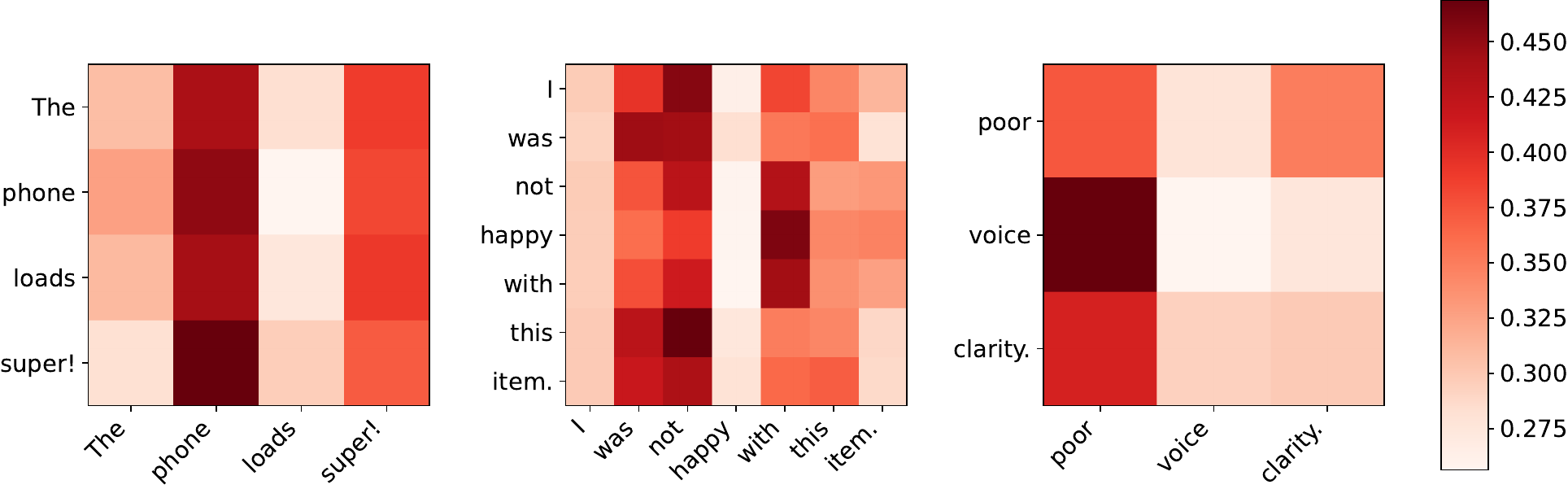}
	\caption{Heat maps of quantum self-attention.}
	\label{fig:Heat maps}
\end{figure}

To further demonstrate the effectiveness of our approach, we visualized the quantum self-attention mechanisms through heat maps, as shown in Fig.\,\ref{fig:Heat maps},

\subsection{Noise robustness}
In the practical application of quantum computing, the impact of quantum noise is a significant factor, as NISQ is sensitive to the environment and susceptible to noise interference. Our investigation consisted of two experimental parts: first, applying single-qubit noise channels (including depolarizing, amplitude damping, and phase damping) to the final circuit layer; second, we inserted two-qubit depolarizing channels after each two-qubit gate throughout the circuit, while maintaining the single-qubit depolarizing noise at 0.01.

Depolarizing channel (D) causes a qubit to depolarize with probability $p$. For a single qubit, it is replaced by the completely mixed state $I/2$, and remains unchanged with probability 1-$p$. The single-qubit depolarizing channel can be represented as the following density matrix mapping:
\begin{equation}
\varepsilon_{\text{D}}(\rho)=(1-p)\rho+\frac{p}{3}(X\rho X+Y\rho Y+Z\rho Z),
\end{equation}
the two-qubit depolarizing channel: 
\begin{equation}
    \varepsilon_{D}(\rho) = (1 - p)\rho + \frac{p}{15} \sum_{\substack{P, Q \in \{I, X, Y, Z\}, \\ (P, Q) \neq (I, I)}} P \otimes Q \, \rho \, P \otimes Q,
\end{equation}
where $\rho$ is the original density matrix, and $X$, $Y$, $Z$ are Pauli matrices.

Amplitude damping (AD) describes the process of a quantum system losing energy, while phase damping (PD) describes the process of a quantum system losing phase information without losing energy. The noise mapping for a single qubit's density matrix can be uniformly expressed as:
\begin{equation}
\varepsilon_{\mathrm{AD/PD}}(\rho)=E_0\rho E_0^\dagger+E_1\rho E_1^\dagger,
\end{equation}
where for amplitude damping, $E_0=\ket{0}\bra{0}+\sqrt{1-p}\ket{1}\bra{1}$ and $E_1=\sqrt{p}\ket{0}\bra{1}$, and for phase damping, $E_0=\ket{0}\bra{0}+\sqrt{1-p}\ket{1}\bra{1}$ and $E_1=\sqrt{p}\ket{1}\bra{1}$. $E_0$ and $E_1$ represent Kraus operators, and $p$ represents the noise level.

\begin{table*}[ht!]
\caption{Test accuracy of QMSAN-P models under various noise channels.}
\label{table:noise_robustness}
\centering
\resizebox{\textwidth}{!}{%
\begin{tabular}{lccccccccc}
\toprule
\multirow{2}{*}{Noise Type} & \multicolumn{3}{c}{Yelp} & \multicolumn{3}{c}{IMDb} & \multicolumn{3}{c}{Amazon} \\
\cmidrule(lr){2-4} \cmidrule(lr){5-7} \cmidrule(lr){8-10}
 & R-P & CB-P & AA-P & R-P & CB-P & AA-P & R-P & CB-P & AA-P \\
\midrule
D(0.01\_s)       & 84.55$\pm$1.79 & 84.51$\pm$1.38 & 84.77$\pm$1.81 & 84.10$\pm$3.14 & 84.27$\pm$2.16 & 83.89$\pm$3.64 & 86.97$\pm$1.66 & 86.98$\pm$1.53 & 86.40$\pm$1.16 \\
D(0.1\_s)        & 84.49$\pm$1.59 & 84.13$\pm$2.27 & 84.11$\pm$2.35 & 84.30$\pm$1.81 & 84.03$\pm$2.07 & 83.58$\pm$1.58 & 87.02$\pm$0.71 & 86.87$\pm$1.21 & 87.23$\pm$1.40 \\
D(0.2\_s)        & 83.97$\pm$1.74 & 83.76$\pm$3.66 & 84.22$\pm$2.34 & 83.29$\pm$3.04 & 83.80$\pm$1.63 & 83.83$\pm$2.66 & 86.17$\pm$1.53 & 86.05$\pm$1.97 & 86.33$\pm$1.50 \\
AD(0.01\_s)       & 84.53$\pm$2.28 & 83.99$\pm$0.97 & 84.63$\pm$1.39 & 84.01$\pm$3.02 & 84.17$\pm$3.12 & 83.77$\pm$2.42 & 86.42$\pm$1.02 & 87.34$\pm$1.17 & 87.21$\pm$1.81 \\
AD(0.1\_s)        & 84.05$\pm$2.30 & 84.15$\pm$2.01 & 84.54$\pm$1.87 & 84.09$\pm$2.59 & 83.98$\pm$3.89 & 83.69$\pm$1.69 & 86.30$\pm$0.51 & 87.29$\pm$0.81 & 86.99$\pm$2.25 \\
AD(0.2\_s)        & 83.87$\pm$1.40 & 83.67$\pm$1.71 & 84.20$\pm$1.86 & 83.95$\pm$2.54 & 84.07$\pm$2.70 & 83.91$\pm$1.71 & 86.33$\pm$2.29 & 86.23$\pm$1.63 & 87.30$\pm$1.72 \\
PD(0.01\_s)       & 84.11$\pm$2.56 & 84.60$\pm$1.83 & 83.90$\pm$1.53 & 84.02$\pm$2.79 & 84.02$\pm$2.79 & 84.05$\pm$2.43 & 87.02$\pm$0.95 & 86.78$\pm$0.75 & 87.05$\pm$1.05 \\
PD(0.1\_s)        & 84.35$\pm$2.73 & 84.22$\pm$2.82 & 83.42$\pm$1.32 & 84.22$\pm$2.73 & 83.93$\pm$2.08 & 83.94$\pm$3.22 & 86.37$\pm$1.96 & 86.32$\pm$1.96 & 86.86$\pm$2.25 \\
PD(0.2\_s)        & 84.33$\pm$1.44 & 84.46$\pm$1.24 & 83.85$\pm$1.33 & 83.66$\pm$2.58 & 84.11$\pm$2.99 & 83.71$\pm$4.06 & 86.91$\pm$2.15 & 86.83$\pm$1.69 & 86.89$\pm$2.36 \\ 
D(0.01\_s+0.01\_t) & 82.57$\pm$1.93 & 82.04$\pm$2.61 & 82.60$\pm$1.55 & 82.58$\pm$2.91 & 82.72$\pm$3.07 & 82.84$\pm$2.35 & 85.92$\pm$1.88 & 85.01$\pm$1.79 & 85.62$\pm$1.24 \\
D(0.01\_s+0.05\_t) & 81.10$\pm$2.18 & 80.11$\pm$1.36 & 81.03$\pm$1.92 & 81.32$\pm$2.14 & 81.35$\pm$2.23 & 81.51$\pm$2.02 & 83.81$\pm$2.24 & 82.93$\pm$2.51 & 84.15$\pm$2.54 \\
\bottomrule
\multicolumn{10}{l}{\textbf{Note} — The noise types used in the experiments are D (Depolarizing), AD (Amplitude damping), and PD (Phase damping).} \\
\multicolumn{10}{l}{The noise notation uses the format \texttt{NoiseType(NoiseLevel\_s + NoiseLevel\_t)}, where \_s refers to single-qubit noise} \\
\multicolumn{10}{l}{and \_t refers to two-qubit noise.}

\end{tabular}
}
\end{table*}

Analyzing the impact of single-qubit noise on the three models, we observe that For single-qubit noise, the test accuracy drop ranges from 0.19\% to 1.48\%. When combining single-qubit depolarizing channels at a 0.01 level with two-qubit depolarizing channels, the accuracy drop ranges from 1.45\% to 4.71\%.

This robust performance against single-qubit noise can be attributed to the variational quantum algorithms (VQAs) employed in our model \cite{fontana2021evaluating,sharma2020noise}. As demonstrated by Ref.\,\cite{fontana2021evaluating}, VQAs mitigate the effects of noise by adapting the optimized parameters. The variational nature of VQAs enables the circuit to adjust its parameters during training in noisy environments, effectively reducing the impact of noise. Furthermore, their work suggests that circuits containing redundant parameterized gates exhibit enhanced resilience to noise. This over parameterization provides a larger optimization landscape, allowing the algorithm to find noise-robust parameter configurations.

Two-qubit noise channels shows a stronger impact on quantum systems. This may be due to two-qubit noise channels simultaneously affect the quantum states of both involved qubits, leading to a more substantial perturbation of the encoded information compared to single-qubit noise. Moreover, two-qubit noise directly reduces entanglement between qubits, which is crucial for quantum correlations. This leads to a more significant performance decline compared to the impact of single-qubit noise.

\section{Conclusion}
\label{section 7}
This paper introduces the Quantum Mixed-State Self-Attention Network (QMSAN), a novel approach that combines quantum computing principles with self-attention mechanisms to enhance natural language processing tasks. Our model leverages the unique properties of quantum systems, particularly similarity computation using mixed states, to improve the efficiency and effectiveness of attention computations. We developed an innovative quantum positional encoding scheme that incorporates positional information directly into the quantum circuit through fixed gates. This advancement enhances the model's ability to capture sequence information without requiring additional qubit resources. Also, our simulation experiments demonstrate that QMSAN exhibits a degree of noise resilience, indicating its potential for implementation on near-term quantum devices.

Looking to the future, we expect more research into quantum machine learning models, with the goal of creating entirely quantum-based attention networks that make the most of quantum computing's special features. As the field of quantum machine learning progresses, we hope our research can add more value and make meaningful contributions to the potential role of quantum computing in improving language processing methods.

\appendix
\section{Analysis of Expressivity and Entangling Capability in Three Quantum Circuit Configurations}
\label{app1}

In our paper, the entangling capability of a circuit is defined as the average Meyer-Wallach entanglement \cite{sim2019expressibility} of its output states:
\begin{equation}
\text{Ent} = \frac{1}{|S|} \sum_{\theta_i\in S} Q(|\psi_{\theta_i}\rangle),
\end{equation}
where $S = {\theta_i}$ is a set of sampled circuit parameter vectors, and $Q$ is the Meyer-Wallach measure. For an n-qubit system, $Q$ is computed using:
\begin{equation}
Q(|\psi\rangle) = \frac{4}{n} \sum_{j=1}^n D(\iota_j(0)|\psi\rangle, \iota_j(1)|\psi\rangle),
\end{equation}
where $\iota_j(b)$ is a linear mapping on the computational basis:
\begin{equation}
\iota_j(b) |b_1...b_n\rangle = \delta_{bb_j} |b_1...\hat{b}_j...b_n\rangle,
\end{equation}
The symbol $\hat{}$ indicates the absence of the $j$-th qubit. The function $D$ measures the generalized distance between quantum states:
\begin{equation}
D(|u\rangle, |v\rangle) = \frac{1}{2} \sum_{i,j} |u_i v_j - u_j v_i|^2.
\end{equation}

\begin{figure*}[htbp]
    \centering
    \subfigure[Expressivity]{
        \begin{minipage}[b]{0.45\textwidth}
            \includegraphics[width=\linewidth]{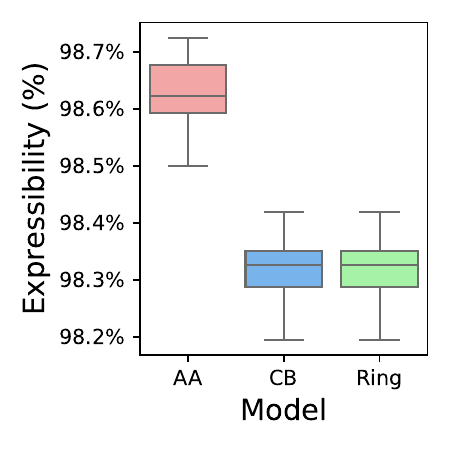} 
        \end{minipage}
        \label{fig:grid_4figs_1cap_4subcap_1}
    }
    \subfigure[Entangling Capability]{
        \begin{minipage}[b]{0.45\textwidth}
            \includegraphics[width=\linewidth]{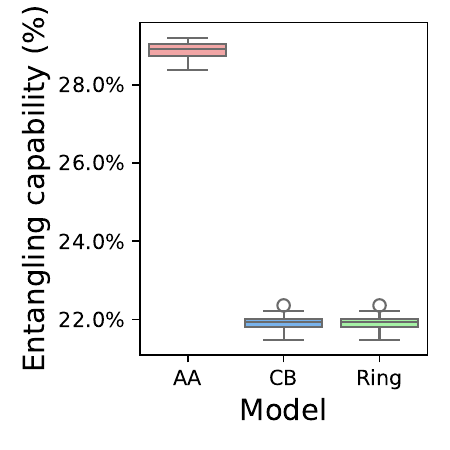}
        \end{minipage}
        \label{fig:grid_4figs_1cap_4subcap_2}
    }
    \caption{Expressivity and Entangling Capability of Three Circuit Configurations}
    \label{fig:Expressivity and Entangling Capability}
\end{figure*}

We define the expressivity measure as Expr=1-MMD, where MMD \cite{ding2022evaluating} is calculated between the output distribution of the quantum circuit and a uniform distribution. The MMD is estimated using:
\begin{equation}
\text{MMD}(P,Q) \approx \frac{1}{N^2} \left| \sum_{i,j=1}^N k(X_i,X_j) + k(Y_i,Y_j) - 2k(X_i,Y_j) \right|,
\end{equation}
where $X_i$ and $Y_i$ are samples from the circuit output distribution $P$ and the uniform distribution $Q$, respectively. The function $k(x,y)$ is a Gaussian kernel:
\begin{equation}
k(x,y) = e^{-\frac{|x-y|^2}{2\sigma^2}},
\end{equation}
where $\sigma$ is a hyperparameter set to 0.01 in our calculations to optimize prediction performance.

To evaluate the expressivity and entanglement capabilities of our three circuit configurations (All-to-All, Circuit-Block, and Ring), we conducted a series of experiments using 4-qubit circuits. We performed 20 random experiments, each sampling 10,000 instances. The results are visualized using box plots in Fig.\,\ref{fig:Expressivity and Entangling Capability}.

The experimental outcomes reveal that the All-to-all (AA) circuit configuration demonstrates superior expressivity compared to both the Circuit-block (CB) and Ring configurations. Interestingly, despite their different architectures, the CB and Ring configurations exhibit similar levels of expressivity and entanglement capability.

\section*{Acknowledgments}
The present work is supported by Education and Scientific Research Project for Young and Middle-aged Teachers of Fujian Province, China (Grant numbers JAT200499) and the Science and Technology Development Fund, Macau SAR (Grant numbers 0093/2022/A2, 0076/2022/A2, and 0008/2022/AGJ)

\bibliographystyle{elsarticle-num}
\bibliography{references}
\end{document}